*Research Article*

# Cores of Cooperative Games in Information Theory

**Mokshay Madiman**

*Department of Statistics, Yale University, 24 Hillhouse Avenue, New Haven, CT 06511, USA*

Correspondence should be addressed to Mokshay Madiman, mokshay.madiman@yale.edu

Cores of cooperative games are ubiquitous in information theory and arise most frequently in the characterization of fundamental limits in various scenarios involving multiple users. Examples include classical settings in network information theory such as Slepian-Wolf source coding and multiple access channels, classical settings in statistics such as robust hypothesis testing, and new settings at the intersection of networking and statistics such as distributed estimation problems for sensor networks. Cooperative game theory allows one to understand aspects of all these problems from a fresh and unifying perspective that treats users as players in a game, sometimes leading to new insights. At the heart of these analyses are fundamental dualities that have been long studied in the context of cooperative games; for information theoretic purposes, these are dualities between information inequalities on the one hand and properties of rate, capacity, or other resource allocation regions on the other.

## 1. INTRODUCTION

A central problem in information theory is the determination of rate regions in data compression problems and that of capacity regions in communication problems. Although single-letter characterizations of these regions were given for lossless data compression of one source and for communication from one transmitter to one receiver by Shannon himself, more elaborate scenarios involving data compression from many correlated sources or communication between a network of users remain of great theoretical and practical interest, with many key problems remaining open. In these multiuser scenarios, rate and capacity regions are subsets of some Euclidean space whose dimension depends on the number of users. The search for an "optimal" rate point is no longer trivial, even if the rate region is known, because of the fact that there is no natural total ordering on points of Euclidean space. Indeed, it is important to ask in the first place what optimality means in the multiuser context—typical criteria for optimality, depending on the scenario of interest, would derive from considerations of fairness, net efficiency, extraneous costs, or robustness to various kinds of network failures.

Our primary goal in this paper is to point out that notions from cooperative game theory arise in a very natural way in connection with the study of rate and capacity regions for several important problems. Examples of these problems include Slepian-Wolf source coding, multiple access channels, and certain distributed estimation problems for sensor networks. Using notions from cooperative game theory, certain properties of the rate regions follow from appropriate information inequalities. In the case of Slepian-Wolf coding and multiple access channels, these results are very well known; perhaps some of the interpretations are unusual, but the experts will not find them surprising. In the case of the distributed estimation setting, the results are recent and the interpretation is new. We supplement the analysis of these rate regions by pointing out that the classical capacity-based theory of composite hypothesis testing pioneered by Huber and Strassen also has a game-theoretic interpretation, but in terms of games with an uncountable infinity of players. Since most of our results concern new interpretations of known facts, we label them as Translations.

The paper is organized as follows. In Section 2, some basic facts from the theory of cooperative games are reviewed. Section 3 treats using the game-theoretic framework the distributed compression problem solved by Slepian and Wolf. The extreme points of the Slepian-Wolf rate region are interpreted in terms of robustness to certain kinds of network failures, and allocations of rates to users that are "fair" or "tolerable" are also discussed. Section 4 considers various classes of multiple access channels. An interesting special case is the Gaussian multiple access channel, where the game associated with the standard setting has significantly nicer structure than the game studied by La and Anantharam [1] associated with an arbitrarily varying setting. Section 5 describes a model for distributed estimation using sensor networks and studies a game associated with allocation of risks for this model. Section 6 looks at various games involving the entropies and entropy powers of sums. These do not seem to have an operational interpretation but are related to recently developed information inequalities. Section 7 discusses connections of the game-theoretic framework with the theory of robust hypothesis testing. Finally, Section 8 contains some concluding remarks.

## 2. A REVIEW OF COOPERATIVE GAME THEORY

The theory of cooperative games is classical in the economics and game theory literature and has been extensively developed. The basic setting of such a game consists of $n$ players, who can form arbitrary coalitions $\mathbf{s} \subset [n]$, where $[n]$ denotes the set $\{1, 2, \ldots, n\}$ of players. A game is specified by the set $[n]$ of players, and a value function $v: 2^{[n]} \to \mathbb{R}_+$, where $\mathbb{R}_+$ is the nonnegative real numbers, and it is always assumed that $v(\phi) = 0$. The value of a coalition $\mathbf{s}$ is equal to $v(\mathbf{s})$.

We will usually interpret the cooperative game (in its standard form) as the setting for a cost allocation problem. Suppose that player $i$ contributes an amount of $t_i$. Since the game is assumed to involve (linearly) transferable utility, the cumulative cost to the players in the coalition $\mathbf{s}$ is simply $\sum_{i \in \mathbf{s}} t_i$. Since each coalition must pay its due of $v(s)$, the individual costs $t_i$ must satisfy $\sum_{i \in \mathbf{s}} t_i \geq v(\mathbf{s})$ for every $\mathbf{s} \subset [n]$. This set of cost vectors, namely

$$A(v) = \left\{ t \in \mathbb{R}_+^n : \sum_{i \in \mathbf{s}} t_i \geq v(\mathbf{s}) \text{ for each } \mathbf{s} \subset [n] \right\}, \quad (1)$$

is the set of *aspirations* of the game, in the sense that this set defines what the players can aspire to. The goal of the game is to minimize social cost, that is, the total sum of the costs $\sum_{i \in [n]} t_i$. Clearly this minimum is achieved when $\sum_{i \in [n]} t_i = v([n])$. This leads to the definition of the core of a game.

*Definition 1.* The core of a game $v$ is the set of aspiration vectors $t \in A(v)$ such that $\sum_{i \in [n]} t_i = v([n])$.

One may think of the core of an arbitrary game as the intersection of the set of aspirations $A(v)$ and the "efficiency hyperplane":

$$F(v) = \left\{ t \in \mathbb{R}^n : \sum_{i \in [n]} t_i = v([n]) \right\}. \quad (2)$$

The core can be equivalently defined as the set of undominated imputations; see, for example, Owen's book [2] for this approach, and a proof of the equivalence. In this paper, we will not consider the question of where the value function of a game comes from but rather take the value function as given and study the corresponding game using structural results from game theory. However, in the original economic interpretation, one should think of $v(\mathbf{s})$ as the amount of utility that the members of $\mathbf{s}$ can obtain from the game whatever the remaining players may do. Then, one can interpret $t_i$ as the payoff to the $i$th player and $v(\mathbf{s})$ as the minimum net payoff to the members of the coalition $\mathbf{s}$ that they will accept. This gives the aspiration set a slightly different interpretation. Indeed, the aspiration set can be thought of as the set of payoff vectors to players that no coalition would block as being inadequate. For the purposes of this paper, one may think of a cooperative game either in terms of payoffs as discussed in this paragraph or in terms of cost allocation as described earlier.

A pathbreaking result in the theory of transferable utility games was the Bondareva-Shapley theorem characterizing whether the core of the game is empty. First, we need to define the notion of a balanced game.

*Definition 2.* Given a collection $\mathcal{C}$ of subsets of $[n]$, a function $\alpha : \mathcal{C} \to \mathbb{R}_+$ is a *fractional partition* if for each $i \in [n]$, we have $\sum_{\mathbf{s} \in \mathcal{C} : i \in S} \alpha(\mathbf{s}) = 1$. A game is *balanced* if

$$v([n]) \geq \sum_{\mathbf{s} \in \mathcal{C}} \alpha(\mathbf{s}) v(\mathbf{s}) \quad (3)$$

for any fractional partition $\alpha$ for any collection $\mathcal{C}$.

Actually, to check that a game is balanced, one does not need to show the inequality (3) for all fractional partitions for all collections $\mathcal{C}$. It is sufficient to check (3) for "minimal balanced collections" (and these collections turn out to yield a unique fractional partition). Details may be found, for example, in Owen [2].

We now state the Bondareva-Shapley theorem [3, 4].

*Fact 1.* The core of a game is nonempty if and only if the game is balanced.

*Proof.* Consider the linear program:

$$\begin{aligned}
\text{Maximize } & \sum_{\mathbf{s} \subset [n]} \alpha(\mathbf{s}) v(\mathbf{s}), \\
\text{subject to } & \alpha(\mathbf{s}) \geq 0 \quad \text{for each } \mathbf{s} \subset [n], \\
& \sum_{\mathbf{s} \subset [n], \mathbf{s} \ni j} \alpha(\mathbf{s}) = 1 \quad \text{for each } j \in [n].
\end{aligned} \quad (4)$$

The dual problem is easily obtained

$$\begin{aligned}
\text{Minimize } & \sum_{j \in [n]} t_j, \\
\text{subject to } & \sum_{j \in \mathbf{s}} t_j \geq v(\mathbf{s}) \quad \text{for each } \mathbf{s} \subset [n].
\end{aligned} \quad (5)$$

If $p^*$ and $d^*$ denote the primal and dual optimal values, duality theory tells us that $p^* = d^*$. Also, the game being balanced means $p^* \leq v([n])$, while the core being nonempty means that $d^* \leq v([n])$. (Note that by setting $\alpha(\mathbf{s}) = 0$ for some subsets $\mathbf{s} \subset [n]$, fractional partitions using arbitrary collections of sets can be thought of as fractional partitions using the full power set $2^{[n]}$.) Thus, the game having a nonempty core is equivalent to its being balanced. □

An important class of games is that of convex games.

*Definition 3.* A game is *convex* if

$$v(\mathbf{s} \cup \mathbf{t}) + v(\mathbf{s} \cap \mathbf{t}) \geq v(\mathbf{s}) + v(\mathbf{t}) \qquad (6)$$

for any sets $\mathbf{s}$ and $\mathbf{t}$. (In this case, the set function $v$ is also said to be supermodular.)

The connection between convexity and balancedness goes back to Shapley.

*Fact 2.* A convex game is balanced and has nonempty core; the converse need not hold.

*Proof.* Shapley [5] showed that convex games have nonempty core, hence they must be balanced by Fact 1. A direct proof by induction of the fact that convexity implies fractional superadditivity inequalities (which include balancedness) is given in [6]. □

Incidentally, Maschler et al. [7] (cf., Edmonds [8]) noticed that the dimension of the core of a convex game was determined by the decomposability of the game, which is a measure of how much "additivity" (as opposed to the kind of superadditivity imposed by convexity) there is in the value function of the game.

There are various alternative characterizations of convex games that are of interest. For any game $v$ and any ordering (permutation) $\sigma = (i_1, \ldots, i_n)$ on $[n]$, the marginal worth vector $m^\sigma(v) \in \mathbb{R}^n$ is defined by

$$m_{i_k}^\sigma(v) = v(\{i_1, \ldots, i_k\}) - v(\{i_1, \ldots, i_{k-1}\}) \qquad (7)$$

for each $k > 1$, and $m_{i_1}^\sigma(v) = v(\{i_1\})$. The convex hull of all the marginal vectors is called the *Weber set*. Weber [9] showed that the Weber set of any game contains its core. The *Shapley-Ichiishi theorem* [5, 10] says that the Weber set is identical to the core if and only if the game is convex. In particular, the extreme points of the core of a convex game are precisely the marginal vectors.

This characterization of convex games is obviously useful from an optimization point of view, as studied deeply by Edmonds [8] in the closely related theory of polymatroids. Indeed, polymatroids (strictly speaking, contrapolymatroids) may simply be thought of as the aspiration sets of convex games. Note that in the presence of the convexity condition, the assumption that $v$ takes only nonnegative values is equivalent to the nondecreasing condition $v(\mathbf{s}) \leq v(\mathbf{t})$ if $\mathbf{s} \subset \mathbf{t}$. Since a linear program is solved at extreme points, the results of Edmonds (stated in the language of polymatroids) and Shapley (stated in the language of convex games) imply that any linear function defined on the core of a convex game (or the dominant face of a polymatroid) must be extremized at a marginal vector. Edmonds [8] uses this to develop greedy methods for such optimization problems. Historically speaking, the two parallel theories of polymatroids and convex games were developed around the same time in the mid-1960s with awareness of and stimulated by each other (as evidenced by a footnote in [5]); however, in information theory, this parallelism does not seem to be part of the folklore, and the game interpretation of rate or capacity regions has only been used to the author's knowledge in the important paper of La and Anantharam [1].

The Shapley value of a game $v$ is the centroid of the marginal vectors:

$$\phi[v] = \frac{1}{n!} \sum_{\sigma \in S_n} m^\sigma, \qquad (8)$$

where $S_n$ is the symmetric group consisting of all permutations. As shown by Shapley [11], its components are given by

$$\phi_i[v] = \sum_{\mathbf{s} \ni i} \frac{(|\mathbf{s}| - 1)!(n - |\mathbf{s}|)!}{n!} \left[ v(\mathbf{s}) - v(\mathbf{s} \setminus \{i\}) \right], \qquad (9)$$

and it is the unique vector satisfying the following axioms: (a) $\phi$ lies in the efficiency hyperplane $F(v)$, (b) it is invariant under permutation of players, and (c) if $u$ and $v$ are two games, then $\phi[u + v] = \phi[u] + \phi[v]$. Clearly, the Shapley value gives one possible formalization of the notion of a "fair allocation" to the players in the game.

*Fact 3.* For a convex game, the Shapley value is in the core.

*Proof.* As pointed out by Shapley [5], this simply follows from the representation of the Shapley value as a convex combination of marginal vectors and the fact that the core of a convex game contains its Weber set. □

For a cooperative game, convexity is quite a strong property. It implies, in particular, both that the game is exact and that it has a large core; we describe these notions below.

If $\sum_{i \in \mathbf{s}} y_i \geq v(\mathbf{s})$ for each $\mathbf{s}$, does there exist $x$ in the core such that $x \leq y$ (component-wise)? If so, the core is said to be *large*. Sharkey [12] showed that not all balanced games have large cores, and that not all games with large cores are convex. However, [12] also showed the following fact.

*Fact 4.* A convex game has a large core.

A game with value function $v$ is said to be *exact* if for every set $\mathbf{s} \subset [n]$, there exists a cost vector $t$ in the core of the game such that

$$\sum_{i \in \mathbf{s}} t_i = v(\mathbf{s}). \qquad (10)$$

Since for any point in the core, the net cost to the members of $\mathbf{s}$ is at least $v(\mathbf{s})$, a game is exact if and only if

$$v(\mathbf{s}) = \min \left\{ \sum_{i \in \mathbf{s}} t_i : t \text{ is in the core of } v \right\}. \qquad (11)$$

The exactness and large core properties are not comparable (counterexamples can be found in [12] and Biswas et al. [13]). However, Schmeidler [14] showed the following fact.

*Fact 5.* A convex game is exact.

Interestingly, Rabie [15] showed that the Shapley value of an exact game need not be in its core.

One may define, in an exactly complementary way to the above development, cooperative games that deal with *resource allocation* rather than cost allocation. The set of aspirations for a resource allocation game is

$$A(v) = \left\{ t \in \mathbb{R}^n_+ : \sum_{i \in \mathbf{s}} t_i \leq v(\mathbf{s}) \text{ for each } \mathbf{s} \subset [n] \right\}, \quad (12)$$

and the core is the intersection of this set with the efficiency hyperplane $F(v)$ defined in (2), which represents the maximum achievable resource for the grand coalition of all players, and thus a public good. A resource allocation game is concave if

$$v(\mathbf{s} \cup \mathbf{t}) + v(\mathbf{s} \cap \mathbf{t}) \leq v(\mathbf{s}) + v(\mathbf{t}) \quad (13)$$

for any sets $\mathbf{s}$ and $\mathbf{t}$. The concavity of a game can be thought of as the "decreasing marginal returns" property of the value function, which is well motivated by economics.

One can easily formulate equivalent versions of Facts 1, 2, 3, 4, and 5 for resource allocation games. For instance, the analogue of Fact 1 is that the core of a resource allocation game is nonempty if and only if

$$v([n]) \leq \sum_{S \in \mathcal{C}} \alpha(s) v(s) \quad (14)$$

for each fractional partition $\alpha$ for any collection of subsets $\mathcal{C}$ (we call this property also balancedness, with some slight abuse of terminology). This follows from the fact that the duality used to prove Fact 1 remains unchanged if we simultaneously change the signs of $\{t_i\}$ and $v$, and reverse relevant inequalities.

Notions from cooperative game theory also appear in the more recently developed theory of combinatorial auctions. In combinatorial auction theory, the interpretation is slightly different, but it remains an economic interpretation, and so we discuss it briefly to prepare the ground for some additional insights that we will obtain from it. Consider a resource allocation game $v: 2^{[n]} \to \mathbb{R}$, where $[n]$ indexes the items available on auction. Think of $v(\mathbf{s})$ as the amount that a bidder in an auction is willing to pay for the particular bundle of items indexed by $\mathbf{s}$. In designing the rules of an auction, one has to take into account all the received bids, represented by a number of such set functions or "valuations" $v$. The auction design then determines how to make an allocation of items to bidders, and computational concerns often play a major role.

We wish to highlight a fact that has emerged from combinatorial auction theory; first we need a definition introduced by Lehmann et al. [16].

*Definition 4.* A set function $v$ is additive if there exist non-negative real numbers $t_1, \ldots, t_n$ such that $v(\mathbf{s}) = \sum_{i \in \mathbf{s}} t_i$ for each $\mathbf{s} \subset [n]$. A set function $v$ is *XOS*, if there are additive value functions $v_1, \ldots, v_M$ for some positive integer $M$ such that

$$v(\mathbf{s}) = \max_{j \in [M]} v_j(\mathbf{s}). \quad (15)$$

The terminology XOS emerged as an abbreviation for "XOR of OR of singletons" and was motivated by the need to represent value functions efficiently (without storing all $2^n - 1$ values) in the computer science literature. Feige [17] proves the following fact, by a modification of the argument for the Bondareva-Shapley theorem.

*Fact 6.* A game has an XOS value function if and only if the game is balanced.

By analogy with the definition of exactness for cost allocation games, a resource allocation game is exact if and only if

$$v(\mathbf{s}) = \max \left\{ \sum_{i \in \mathbf{s}} t_i : t \text{ is in the core of } v \right\}. \quad (16)$$

In other words, for an exact game, the additive value functions in the XOS representation of the game can be taken to be those corresponding to the elements of the core (if we allow maximizing over a potentially infinite set of additive value functions).

Some of the concepts elaborated in this section can be extended to games with infinitely many players, although many new technicalities arise. Indeed, there is a whole theory of so-called "nonatomic games" in the economics literature. This is briefly alluded to in Section 7, where we discuss an example of an infinite game.

## 3. THE SLEPIAN-WOLF GAME

The Slepian-Wolf problem refers to the problem of losslessly compressing data from two correlated sources in a distributed manner. Let $p(x_1, \ldots, x_n)$ denote the joint probability mass function of the sources $(X_1, \ldots, X_n) = X_{[n]}$, which take values in discrete alphabets. When the sources are coded in a centralized manner, any rate $R > H(X_{[n]})$ (in bits per symbol) is sufficient, where $H$ denotes the joint entropy, that is, $H(X_{[n]}) = E[-\log p(x_1, \ldots, x_n)]$. What rates are achievable when the sources must be coded separately? This problem was solved for i.i.d sources by Slepian and Wolf [18] and extended to jointly ergodic sources using a binning argument by Cover [19].

*Fact 7.* Correlated sources $(X_1, \ldots, X_n)$ can be described separately at rates $(R_1, \ldots, R_n)$ and recovered with arbitrarily low error probability by a common decoder if and only if

$$\sum_{i \in \mathbf{s}} R_i \geq H(X_\mathbf{s} \mid X_{\mathbf{s}^c}) =: v_{\text{SW}}(\mathbf{s}) \quad (17)$$

for each $\mathbf{s} \subset [n]$. In other words, the Slepian-Wolf rate region is the set of aspirations of the cooperative game $v_{\text{SW}}$, which we call the Slepian-Wolf game.

A key consequence is that using only knowledge of the joint distribution of the data, one can achieve a compression rate equal to the joint entropy of the users (i.e., there is no loss from the incapability to communicate). However, this is not automatic from the characterization of the rate region above; one needs to check that the Slepian-Wolf game is balanced. The balancedness of the Slepian-Wolf game is precisely the content of the lower bound in the following inequality of Madiman and Tetali [6]: for any fractional partition $\alpha$ using $\mathcal{C}$,

$$\sum_{S \in \mathcal{C}} \alpha(\mathbf{s}) H(X_{\mathbf{s}} \mid X_{\mathbf{s}^c}) \leq H(X_{[n]}) \leq \sum_{S \in \mathcal{C}} \alpha(\mathbf{s}) H(X_{\mathbf{s}}). \quad (18)$$

This weak fractional form of the joint entropy inequalities in [6] coupled with Fact 1 proves that the joint entropy is an achievable sum rate even for distributed compression. In fact, the Slepian-Wolf game is much nicer.

*Translation 1.* The Slepian-Wolf game is a convex game.

*Proof.* To show that the Slepian-Wolf game is convex, we need to show that $v_{\text{SW}}(\mathbf{s}) = H(X_{\mathbf{s}} \mid X_{\mathbf{s}^c})$ is supermodular. This fact was first explicitly pointed out by Fujishige [20]. □

By applying Fact 2, the core is nonempty since the game is convex, which means that there exists a rate point satisfying

$$\sum_{i \in [n]} R_i = v_{\text{SW}}([n]) = H(X_{[n]}). \quad (19)$$

This recovers the fact that a sum rate of $H(X_{[n]})$ is achievable. Note that, combined with Fact 1, this observation in turn gives an immediate proof of the inequality (18).

We now look at how robust this situation is to network degradation because some users drop out. First note that by Fact 5, the Slepian-Wolf game is exact. Hence, for any subset $\mathbf{s}$ of users, there exists a vector $R = (R_1, \ldots, R_n)$ that is sum-rate optimal for the grand coalition of all users, which is also sum-rate optimal for the users in $\mathbf{s}$, that is, $\sum_{i \in \mathbf{s}} R_i = v_{\text{SW}}(\mathbf{s})$. However, in general, it is not possible to find a rate vector that is simultaneously sum-rate optimal for multiple proper subsets of users. Below, we observe that finding such a rate vector is possible if the subsets of interest arise from users potentially dropping out in a certain order.

*Translation 2* (Robust Slepian-Wolf coding). Suppose the users can only drop out in a certain order, which without loss of generality we can take to be the natural decreasing order on $[n]$ (i.e., we assume that the first user to potentially drop out would be user $n$, followed by user $n-1$, etc.). Then, there exists a rate point for Slepian-Wolf coding which is feasible and optimal irrespective of the number of users that have dropped out.

*Proof.* The solution to this problem is related to a modified Slepian-Wolf game, given by the utility function:

$$\bar{v}_{\text{SW}}(\mathbf{s}) = H(X_{\mathbf{s}} \mid X_{\mathbf{s}^c \setminus > \mathbf{s}}), \quad (20)$$

where $> \mathbf{s} = \{i \in [n] : i > j \text{ for every } j \in \mathbf{s}\}$. Indeed, if this game is shown to have a nonempty core, then there exists a rate point which is simultaneously in the Slepian-Wolf rate region of every $[k]$, for $k \in [n]$. However, the nonemptiness of the core is equivalent to the balancedness of $\bar{v}_{\text{SW}}$, which follows from the inequality

$$H(X_{[n]}) \geq \sum_{S \in \mathcal{C}} \alpha(\mathbf{s}) H(X_{\mathbf{s}} \mid X_{\mathbf{s}^c \setminus > \mathbf{s}}), \quad (21)$$

where $\alpha$ is any fractional partition using $\mathcal{C}$, which was proved by Madiman and Tetali [6]. To see that the core of this modified game actually contains an optimal point (i.e., a point in the core of the subgame corresponding to the first $k$ users) for each $k$, simply note that the marginal vector corresponding to the natural order on $[n]$ gives a constructive example. □

The main idea here is known in the literature, although not interpreted or proved in this fashion. Indeed, other interpretations and uses of the extreme points of the Slepian-Wolf rate region are discussed, for example, in Coleman et al. [21], Cristescu et al. [22], and Ramamoorthy [23].

It is interesting to interpret some of the game-theoretic facts described in Section 2 for the Slepian-Wolf game. This is particularly useful when there is no natural ordering on the set of players, but rather our goal is to identify a permutation-invariant (and more generally, a "fair") rate point. By Fact 3, we have the following translation.

*Translation 3.* The Shapley value of the Slepian-Wolf game satisfies the following properties. (a) It is in the core of the Slepian-Wolf game, and hence is sum-rate optimal. (b) It is a fair allocation of compression rates to users because it is permutation-invariant. (c) Suppose an additional set of $n$ sources, independent of the first $n$, is introduced. Suppose the Shapley values of the Slepian-Wolf games for the first set of sources is $\phi_1$, and for the second set of sources is $\phi_2$. If each source from the first set is paired with a distinct source from the second set, then the Shapley value for the Slepian-Wolf game played by the set of pairs is $\phi_1 + \phi_2$. (In other words, the "fair" allocation for the pair can be "fairly" split up among the partners in the pair.)

It is pertinent to note, moreover, that implementing Slepian-Wolf coding at any point in the core is practically implementable. While it has been noticed for some time that one can efficiently construct codebooks that nearly achieve the rates at an extreme point of the core, Coleman et al. [21], building on work of Rimoldi and Urbanke [24] in the multiple access channel setting, show a practical approach to efficient coding for any rate point in the core (based on viewing any such rate point as an extreme point of the core of a Slepian-Wolf game for a larger set of sources).

Fact 4 says that the Slepian-Wolf game has a large core, which may be interpreted as follows.

*Translation 4.* Suppose, for each $i$, $T_i$ is the maximum compression rate that user $i$ is willing to tolerate. A tolerance vector $T = (T_1, \ldots, T_n)$ is said to be feasible if

$$\sum_{i \in \mathbf{s}} T_i \geq v_{\text{SW}}(\mathbf{s}) \quad (22)$$

for each $\mathbf{s} \subset [n]$. Then, for any feasible tolerance vector $T$, it is always possible to find a rate point $R = (R_1, \ldots, R_n)$ in the core so that $R_i \leq T_i$ (i.e., the rate point is tolerable to all users).

## 4. MULTIPLE ACCESS CHANNELS AND GAMES

A multiple access channel (MAC) refers to a channel between multiple independent senders (the data sent by the $i$th sender is typically denoted $X_i$) and one receiver (the received data is typically denoted $Y$). The channel characteristics, defined for each transmission by a probability transition $p(y \mid x_1, \ldots, x_n)$, is assumed to be known. We will further restrict our discussion to the case of memoryless channels, where each transmission is assumed to occur independently according to the channel transition probability.

Even within the class of memoryless multiple access channels, there are several notable special cases of interest. The first is the *discrete* memoryless multiple access channel (DM-MAC), where all random variables take values in possibly different finite alphabets, but the channel transition matrix is otherwise unrestricted. The second is the *Gaussian* memoryless multiple access channel (G-MAC); here each sender has a power constraint $P_i$, and the noise introduced to the superposition of the data from the sources is additive Gaussian noise with variance $N$. In other words,

$$Y = \sum_{i \in [n]} X_i + Z, \qquad (23)$$

where $X_i$ are the independent sources, and $Z$ is a mean-zero, variance $N$ is normal independent of the sources. Note that although the power constraints are an additional wrinkle to the problem compared to the DM-MAC, the G-MAC is in a sense more special because of the strong assumption; it makes on the nature of the channel. A third interesting special case is the *Poisson* memoryless multiple access channel (P-MAC), which models optical communication from many senders to one receiver and operates in continuous time. Here, the channel takes in as inputs data from the $n$ sources in the form of waveforms $X_i(t)$, whose peak powers are constrained by some number $A$; in other words, for each sender $i$, $0 \leq X_i(t) \leq A$. The output of the channel is a Poisson process of rate:

$$\lambda_0 + \sum_{i \in [n]} X_i(t), \qquad (24)$$

where the nonnegative constant $\lambda_0$ represents the rate of a homogeneous Poisson process (noise) called the dark current. For further details, one may consult the references cited below.

The capacity region of the DM-MAC was first found by Ahlswede [25] (see also Liao [26] and Slepian and Wolf [27]). Han [28] developed a clear approach to an even more general problem; he used in a fundamental way the polymatroidal properties of entropic quantities, and thus it is no surprise that the problem is closely connected to cooperative games. Below $I$ denotes mutual information (see, e.g., [29]); for notational convenience, we suppress the dependence of the mutual information on the joint distribution.

*Fact 8.* Let $\mathcal{P}$ be the class of joint distributions on $(X_{[n]}, Y)$ for which the marginal on $X_{[n]}$ is a product distribution, and the conditional distribution of $Y$ given $X_{[n]}$ is fixed by the channel characteristics. For $\mu \in \mathcal{P}$, let $\mathbf{C}_\mu$ be the set of capacity vectors $(C_1, \ldots, C_n)$ satisfying

$$\sum_{i \in \mathbf{s}} C_i \leq I(X_\mathbf{s}; Y \mid X_{\mathbf{s}^c}) \qquad (25)$$

for each $\mathbf{s} \subset [n]$. The capacity region of the $n$-user DM-MAC is the closure of the convex hull of the union $\cup \{\mathbf{C}_\mu : \mu \in \mathcal{P}\}$.

This rate region is more complex than the Slepian-Wolf rate region because it is the closed convex hull of the union of the aspiration sets of many cooperative games, each corresponding to a product distribution on $X_{[n]}$. Yet the analogous result turns out to hold. More specifically, even though the different senders have to code in a distributed manner, a sum capacity can be achieved that may be interpreted as the capacity of a single channel from the combined set of sources (coded together).

*Translation 5.* The DM-MAC capacity region is the union of the aspiration sets of a class of concave games. In particular, a sum capacity of $\sup I(X_{[n]}; Y)$ is achievable, where the supremum is taken over all joint distributions on $(X_{[n]}, Y)$ that lie in $\mathcal{P}$.

*Proof.* Let $\Gamma$ denote the set of all conditional mutual information vectors (in the Euclidean space of dimension $2^n$) corresponding to the discrete distributions on $(X_{[n]}, Y)$ that lie in $\mathcal{P}$. More precisely, corresponding to any joint distribution in $\mathcal{P}$ is a point $\gamma \in \Gamma$ defined by

$$\gamma(\mathbf{s}) = I(X_\mathbf{s}; Y \mid X_{\mathbf{s}^c}) \qquad (26)$$

for each $\mathbf{s} \subset [n]$. Han [28] showed that for any joint distribution in $\mathcal{P}$, $\gamma(\mathbf{s})$ is a submodular set function. In other words, each point $\gamma \in \Gamma$ defines a concave game.

As shown in [28], the DM-MAC capacity region may also be characterized as the union of the aspiration sets of games from $\Gamma^*$, where $\Gamma^*$ is the closure of the convex hull of $\Gamma$. It remains to check that each point in $\Gamma^*$ corresponds to a concave game, and this follows from the easily verifiable facts that a convex combination of concave games is concave, and that a limit of concave games is concave.

For the second assertion, note that for any $\gamma \in \Gamma^*$, a sum capacity of $\gamma([n])$ is achievable by Fact 2 (applied to resource allocation games). Combining this with the above characterization of the capacity region and the fact that $\gamma([n]) = I(X_{[n]}; Y)$ for $\gamma \in \Gamma$ completes the argument. □

We now take up the G-MAC. The additive nature of the G-MAC is reflected in a simpler game-theoretic description of its capacity region.

*Fact 9.* The capacity region of the $n$-user G-MAC is the set of capacity allocations $(C_1, \ldots, C_n)$ that satisfy

$$\sum_{i \in \mathbf{s}} C_i \leq C\left(\frac{\sum_{i \in \mathbf{s}} P_i}{N}\right) =: \nu_g(\mathbf{s}) \qquad (27)$$

for each $\mathbf{s} \subset [n]$, where $C(x) = (1/2)\log(1 + x)$. In other words, the capacity region of the G-MAC is the aspiration set of the game defined by $\nu_g$, which we may call the G-MAC game.

*Translation 6.* The G-MAC game is a concave game. In particular, its core is nonempty, and a sum capacity of $C(\sum_{i\in[n]} P_i/N)$ is achievable.

As in the previous section, we may ask whether this is robust to network degradation in the form of users dropping out, at least in some order; the answer is obtained in an exactly analogous fashion.

*Translation 7* (Robust coding for the G-MAC). Suppose the senders can only drop out in a certain order, which without loss of generality we can take to be the natural decreasing order on $[n]$ (i.e., we assume that the first user to potentially drop out would be sender $n$, followed by sender $n - 1$, etc.). Then, there exists a capacity allocation to senders for the G-MAC which is feasible and optimal irrespective of the number of users that have dropped out.

Furthermore, just as for the Slepian-Wolf game, Fact 4 has an interpretation in terms of tolerance vectors analogous to Translation 4. When there is no natural ordering of senders, Fact 3 suggests that the Shapley value is a good choice of capacity allocation for the G-MAC game. Practical implementation of an arbitrary capacity allocation point in the core is discussed by Rimoldi and Urbanke [24] and Yeh [30].

While the ground for the study of the geometry of the G-MAC capacity region using the theory of polymatroids was laid by Han, such a study and its implications were further developed, and in the more general setting of fading that allows the modeling of wireless channels, by Tse and Hanly [31] (see also [30]). Clearly statements like Translation 7 can be carried over to the more general setting of fading channels by building on the observations made in [31].

La and Anantharam [1] provide an elegant analysis of capacity allocation for a different Gaussian MAC model using cooperative game theoretic ideas. We briefly review their results in the context of the preceding discussion.

Consider an Gaussian multiple access channel that is *arbitrarily varying*, in the sense that the users are potentially hostile, aware of each others' codebooks, and are capable of forming "jamming coalitions". A jamming coalition is a set of users, say $\mathbf{s}^c$, who decide not to communicate but instead get together and jam the channel for the remaining users, who constitute the communicating coalition $\mathbf{s}$. As before, each user has a power constraint; the $i$th sender cannot use power greater than $P_i$ whether it wishes to communicate or jam. It is still a Gaussian MAC because the received signal is the superposition of the inputs provided by all the senders, plus additive Gaussian noise of variance $N$. In [1], the value function $\nu_{LA}$ for the game corresponding to this channel is derived; the value for a coalition $\mathbf{s}$ is the capacity achievable by the users in $\mathbf{s}$ even when the users in $\mathbf{s}^c$ coherently combine to jam the channel.

*Fact 10.* The capacity region of the arbitrarily varying Gaussian MAC with potentially hostile senders is the aspiration set of the La-Anantharam game, defined by

$$\nu_{LA}(\mathbf{s}) := C\left(\frac{P_{\hat{\mathbf{s}}}}{\Lambda_{\mathbf{s}^c} + N}\right), \qquad (28)$$

where $P_{\mathbf{s}} = \sum_{i \in \mathbf{s}} P_i$, $\Lambda_{\mathbf{s}} = [\sum_{i \in \mathbf{s}} \sqrt{P_i}]^2$, and $\hat{\mathbf{s}} = \{i \in \mathbf{s} : P_i \geq \Lambda_{\mathbf{s}^c}\}$.

Note that two things have changed relative to the naive G-MAC game; the power available for transmission (appearing in the numerator of the argument of the $C$ function) is reduced because some senders are rendered incapable of communicating by the jammers, and the noise term (appearing in the denominator) is no longer constant for all coalitions but is augmented by the power of the jammers. This tightening of the aspiration set of the La-Anantharam game versus the G-MAC game causes the concavity property to be lost.

*Translation 8.* The La-Anantharam game is not a concave game, but it has a nonempty core. In particular, a sum capacity of $C(\sum_{i\in[n]} P_i/N)$ is achievable.

*Proof.* La and Anantharam [1] show that the Shapley value need not lie in the core of their game, but they demonstrate the existence of another distinguished point in the core. By the analogue of Fact 3 for resource allocation games, the La-Anantharam game cannot be concave. □

Although [1] shows that the Shapley value may not lie in the core, they demonstrate the existence of a unique capacity point that satisfies three desirable axioms: (a) efficiency, (b) invariance to permutation, and (c) envy-freeness. While the first two are also among the Shapley value axioms, [1] provides justification for envy-freeness as an appropriate axiom from the point of view of applications.

We mention here a natural question that we leave for the reader to ponder: given that the La-Anantharam game is balanced but not concave, is it exact? Note that the fact that the Shapley value does not lie in the core is not incompatible with exactness, as shown by Rabie [15].

Finally, we turn to the P-MAC. Lapidoth and Shamai [32] performed a detailed study of this communication problem and showed in particular that the capacity region when all users have the same peak power constraint is given as the closed convex hull of the union of aspiration sets of certain games, just as in the case of the DM-MAC. As in that case, one may check that the capacity region is in fact the union of aspiration sets of a class of concave games, and in particular, as shown in [32], the maximum throughput that one may hope for is achievable.

Of course, there is much more to the well-developed theory of multiple access channels than the memoryless scenarios (discrete, Gaussian and Poisson) discussed above. For instance, there is much recent work on multiuser channels with memory and also with feedback (see, e.g., Tatikonda [33] for a deep treatment of such problems at the intersection of communication and control). We do not discuss these works further, except to make the observation that things can change considerably in these more general scenarios. Indeed, it is quite conceivable that the appropriate games for these scenarios are not convex or concave, and it is even conceivable that such games may not be balanced, which may mean that there are unexpected limitations to achieving the sum rate or sum capacity that one may hope for at first sight.

## 5. A DISTRIBUTED ESTIMATION GAME

In the nascent theory of distributed estimation using sensor networks, one wishes to characterize the fundamental limits of performing statistical tasks such as parameter estimation using a sensor network and apply such characterizations to problems of cost or resource allocation. We discuss one such question for a toy model for distributed estimation introduced by Madiman et al. [34]. By ignoring communication, computation, and other constraints, this model allows one to study the central question of fundamental *statistical* limits without obfuscation.

The model we consider is as follows. The goal is to estimate a parameter $\theta$, which is some unknown real number. Consider a class of sensors, all of which have estimating $\theta$ as their goal. However, the sensors cannot measure $\theta$ directly; they are immersed in a field of sources (that do not depend on $\theta$ and may be considered as producers of noise for the purposes of estimating $\theta$). More specifically, suppose there are $n$ sources, with each source producing a data sample of size $M$ according to some known probability distribution. Let us say that source $i$ generates $X_{i,1}, \ldots, X_{i,M}$. The class of sensors available corresponds to a class $\mathcal{C}$ of subsets of $[n]$, which indexes the set of sources. Owing to the geographical placement of the sensors or for other reasons, each sensor only sees certain aggregate data; indeed, the sensor corresponding to a subset $\mathbf{s} \subset [n]$, known as the **s**-sensor, only sees at any given time the sum of $\theta$ and the data coming from the sources in the set **s**. In other words, the **s**-sensor has access to the observations $\mathbf{Y_s} = (Y_{\mathbf{s},1}, Y_{\mathbf{s},2}, \ldots, Y_{\mathbf{s},M})$, where

$$Y_{\mathbf{s},j} = \theta + \sum_{i \in \mathbf{s}} X_{i,j}. \tag{29}$$

Clearly, $\theta$ shows up as a common location parameter for the observations seen by any sensor.

From the observations $\mathbf{Y_s}$ that are available to it, the **s**-sensor constructs an estimator $\hat{\theta}_\mathbf{s}(\mathbf{Y_s})$ of the unknown parameter $\theta$. The goodness of an estimator is measured by comparing to the "best possible estimator in the worst case", that is, by comparing the risk of the given estimator with the minimax risk. If the risk is measured in terms of mean squared error, then the minimax risk achievable by the **s**-sensor is

$$r_M(\mathbf{s}) = \min_{\text{all estimators } \hat{\theta}_\mathbf{s}} \max_{\theta} E[\hat{\theta}_\mathbf{s}(\mathbf{Y_s}) - \theta]^2. \tag{30}$$

(For location parameters, Girshick and Savage [35] showed that there exists an estimator that achieves this minimax risk.)

The cost measure of interest in this scenario is error variance. Suppose we can give *variance permissions* $V_i$ for each source, that is, the **s**-sensor is only allowed an unbiased estimator with variance not more than $\sum_{i \in \mathbf{s}} V_i$, or more generally, an estimator with mean squared risk not more than this number. For the variance permission vector $(V_1, \ldots, V_n)$ to be feasible with respect to an arbitrary sensor configuration (i.e., for there to exist an estimator for the **s**-user with worst-case risk bounded by $\sum_{i \in \mathbf{s}} V_i$, for every **s**), we need that

$$\sum_{i \in \mathbf{s}} V_i \geq r_M(\mathbf{s}) \tag{31}$$

for each $\mathbf{s} \subset [n]$. Thus, we have the following fact.

*Fact 11.* The set of feasible variance permission vectors is the aspiration set of the cost allocation game

$$v_{\text{DE}}(\mathbf{s}) := r_M(\mathbf{s}), \tag{32}$$

which we call the distributed estimation game.

The natural question is the following. Is it possible to allot variance permissions in such a way that there is no wasted total variance, that is, $\sum_{i \in [n]} V_i = r_M([n])$, and the allotment is feasible for arbitrary sensor configurations? The affirmative answer is the content of the following result.

*Translation 9.* Assuming that all sources have finite variance, the distributed estimation game is balanced. Consequently, there exists a feasible variance allotment $(V_1, \ldots, V_n)$ to sources in $[n]$ such that the $[n]$-sensor cannot waste any of the variance allotted to it.

*Proof.* The main result of Madiman et al. [34] is the following inequality relating the minimax risks achievable by the **s**-users from the class $\mathcal{C}$ to the minimax risk achievable by the $[n]$-user, that is, one who only sees observations of $\theta$ corrupted by *all* the sources. Under the finite variance assumption, for any sample size $M \geq 1$,

$$r_M([n]) \geq \sum_{\mathbf{s} \in \mathcal{C}} \beta(\mathbf{s}) r_M(\mathbf{s}) \tag{33}$$

holds for any fractional partition $\beta$ using any collection of subsets $\mathcal{C}$. In other words, the game $v_{\text{DE}}$ is balanced. Fact 1 now implies that the core is nonempty, that is, a total variance as low as $r_M([n])$ is achievable. □

Translation 9 implies that the optimal sum of variance permissions that can be achieved in a distributed fashion using a sensor network is the same as the best variance that can be achieved using a single centralized sensor that sees all the sources.

Other interesting questions relating to sensor networks can be answered using the inequality (33). For instance, it suggests that using a sensor configuration corresponding to the class $\mathcal{C}_1$ of all singleton sets is better than using a sensor configuration corresponding to the class $\mathcal{C}_2$ of all sets of size 2. We refer the reader to [34] for details. An interesting open problem is the determination of whether this distributed estimation game has a large core.

## 6. AN ENTROPY POWER GAME

The entropy power of a continuous random vector $X$ is $\mathcal{N}(X) = \exp\{2h(X)/d\}/2\pi e$, where $h$ denotes differential entropy. Entropy power plays a key role in several problems of multiuser information theory, and entropy power inequalities have been key to the determination of some capacity and rate regions. (Such uses of entropy power inequalities may be found, e.g., in Shannon [36], Bergmans [37], Ozarow [38], Costa [39], and Oohama [40].) Furthermore, rate regions for several multiuser problems, as discussed already, involve subset sum constraints. Thus, it is conceivable that there exists an interpretation of the discussion below in terms of a multiuser communication problem, but we do not know of one.

We make the following conjecture.

**Conjecture 1.** *Let $X_1, \ldots, X_n$ be independent $\mathbb{R}^d$-valued random vectors with densities and finite covariance matrices. Suppose the region of interest is the set of points $(R_1, \ldots, R_n) \in \mathbb{R}^n_+$ satisfying*

$$\sum_{j \in \mathbf{s}} R_j \geq \mathcal{N}\left(\sum_{j \in \mathbf{s}} X_j\right) \qquad (34)$$

*for each $\mathbf{s} \subset [n]$. Then, there exists a point in this region such that the total sum $\sum_{j \in [n]} R_j = \mathcal{N}(\sum_{j \in [n]} X_j)$.*

By Fact 1, the following conjecture, implicitly proposed by Madiman and Barron [41], is equivalent.

**Conjecture 2.** *Let $X_1, \ldots, X_n$ be independent $\mathbb{R}^d$-valued random vectors with densities and finite covariance matrices. For any collection $\mathcal{C}$ of subsets of $[n]$, let $\beta$ be a fractional partition. Then,*

$$\mathcal{N}(X_1 + \cdots + X_n) \geq \sum_{\mathbf{s} \in \mathcal{C}} \beta(\mathbf{s}) \mathcal{N}\left(\sum_{j \in \mathbf{s}} X_j\right). \qquad (35)$$

*Equality holds if and only if all the $X_i$ are normal with proportional covariance matrices.*

Note that Conjecture 2 simply states that the "entropy power game" defined by $v_{\text{EP}}(\mathbf{s}) := \mathcal{N}(\sum_{j \in \mathbf{s}} X_j)$ is balanced. Define the *maximum degree* in $\mathcal{C}$ as $r_+ = \max_{i \in [n]} r(i)$, where the *degree* $r(i)$ of $i$ in $\mathcal{C}$ is the number of sets in $\mathcal{C}$ that contain $i$. Madiman and Barron [41] showed that Conjecture 2 is true if $\beta(\mathbf{s})$ is replaced by $1/r_+$, where $r_+$ is the maximum degree in $\mathcal{C}$. When every index $i$ has the same degree, $\beta(\mathbf{s}) = 1/r_+$ is indeed a fractional partition.

The equivalence of Conjectures 1 and 2 serves to underscore the fact that the balancedness inequality of Conjecture 2 may be regarded as a more fundamental property (if true) than the generalized entropy power inequalities in [41] and is therefore worthy of attention. The interested reader may also wish to consult [42], where we give some further evidence towards its validity. Of course, if the entropy power game above turns out to be balanced, a natural next question would be whether it is exact or even convex.

While on the topic of games involving the entropy of sums, it is worth mentioning that much more is known about the game with value function:

$$v_{\text{sum}}(\mathbf{s}) := H\left(\sum_{i \in \mathbf{s}} X_i\right), \qquad (36)$$

where $H$ denotes discrete entropy, and $X_i$ are independent discrete random variables. Indeed, as shown by the author in [42], this game is concave, and in particular, has a nonempty core which is the convex hull of its marginal vectors.

For independent continuous random vectors, the set function

$$v(\mathbf{s}) = h\left(\sum_{i \in \mathbf{s}} X_i\right), \qquad (37)$$

where $h$ denotes differential entropy, is submodular as in the discrete case. However, this set function does *not* define a game; indeed, the appropriate convention for $v(\phi)$ is that $v(\phi) = -\infty$, since the null set corresponds to looking at the differential entropy of a constant (say, zero), which is $-\infty$. Because of the fact that the set function $v$ is not real-valued, the submodularity of $v$ does not imply that it is even subadditive (and thus $v$ certainly does not satisfy the inequalities that define balancedness). On the other hand, if $X$ is a continuous random vector independent of $X_1, \ldots, X_n$, and with differentiall entropy $h(X) = 0$, then the modified set function

$$\bar{v}_{\text{sum}}(\mathbf{s}) = h\left(X + \sum_{s \in \mathcal{C}} X_i\right) \qquad (38)$$

is indeed the value function of a balanced cooperative game; see [42] for details and further discussion.

## 7. GAMES IN COMPOSITE HYPOTHESIS TESTING

Interestingly, similar notions also come up in the study of composite hypothesis testing but in the setting of a cooperative resource allocation game for infinitely many users. Let $(\Omega, \mathcal{A})$ be a Polish space with its Borel $\sigma$-lgebra, and let $\mathcal{M}$ be the space of probability measures on $(\Omega, \mathcal{A})$. We may think of $\Omega$ as a set of infinitely many "microscopic players", namely $\omega \in \Omega$. The allowed coalitions of microscopic users are the Borel sets. For our purposes, we specify an infinite cooperative game using a value function $v : \mathcal{A} \to \mathbb{R}$ that satisfies the following conditions:

(1) $v(\phi) = 0$, and $v(\Omega) = 1$,
(2) $A \subset B \Rightarrow v(A) \leq v(B)$,
(3) $A_n \uparrow A \Rightarrow v(A_n) \uparrow v(A)$,
(4) for closed sets $F_n$ with $F_n \downarrow F$, $v(F_n) \downarrow v(F)$.

The continuity conditions are necessary regularity conditions in the context of infinitely many players. The normalization $v(\Omega) = 1$ (which is also sometimes imposed in the study of finite games) is also useful. In the mathematics literature, a value function satisfying the itemized conditions is called a *capacity*, while in the economics literature, it is a *(0,1)-normalized nonatomic game* (usually additional conditions are imposed for the latter). There are many subtle analytical issues that emerge in the study of capacities and nonatomic games. We avoid these and simply mention some infinite analogues of already stated facts.

For any capacity $v$, one may define the family of probability measures

$$\mathcal{P}_v = \{P \in \mathcal{M} : P(A) \leq v(A) \text{ for each } A \in \mathcal{A}\}. \quad (39)$$

The set $\mathcal{P}_v$ may be thought of as the core of the game $v$. Indeed, the additivity of a measure on disjoint sets is the continuous formulation of the transferable utility assumption that earlier caused us to consider sums of resource allocations, while the restriction to *probability* measures ensures efficiency, that is, the maximum possible allocation for the full set.

Let $\mathcal{P}$ be a family of probability measures on $(\Omega, \mathcal{A})$. The set function

$$v(A) = \sup_{P \in \mathcal{P}} P(A), \quad A \in \mathcal{A}, \quad (40)$$

is called the *upper envelope* of $\mathcal{P}$, and it is a capacity if $\mathcal{P}$ is weakly compact. Note that such upper envelopes are just the analogue of the XOS valuations defined in Section 2 for the finite setting. By an extension of Fact 6 to infinite games, one can deduce that the core $\mathcal{P}_v$ is nonempty when $v$ is the upper envelope game for a weakly compact family $\mathcal{P}$ of probability measures.

We say that the infinite game $v$ is concave if, for all measurable sets **s** and **t**,

$$v(\mathbf{s} \cup \mathbf{t}) + v(\mathbf{s} \cap \mathbf{t}) \leq v(\mathbf{s}) + v(\mathbf{t}). \quad (41)$$

In the mathematics literature, the value function of a concave infinite game is often called a *2-alternating capacity*, following Choquet's seminal work [43]. When $v$ defines a concave infinite game, $\mathcal{P}_v$ is nonempty; this is an analog of Fact 2. Furthermore, by the analog of Fact 5, $v$ is an exact game since it is concave, and as in the remarks after Fact 6, it follows that $v$ is just the upper envelope of $\mathcal{P}_v$.

Concave infinite games $v$ are not just of abstract interest; the families of probability measures $\mathcal{P}_v$ that are their cores include important classes of families such as total variation neighborhoods and contamination neighborhoods, as discussed in the references cited below.

A famous, classical result of Huber and Strassen [44] can be stated in the language of infinite cooperative games. Suppose one wishes to test between the composite hypotheses $\mathcal{P}_u$ and $\mathcal{P}_v$, where $u$ and $v$ define infinite games. The criterion that one wishes to minimize is the decay rate of the probability of one type of error in the worst case (i.e., for the worst pair of sources in the two classes), given that the error probability of the other kind is kept below some small constant; in other words, one is using the minimax criterion in the Neyman-Pearson framework. Note that the selection of a critical region for testing is, in the game language, the selection of a coalition. In the setting of simple hypotheses, the optimal coalition is obtained as the set for which the Radon-Nikodym derivative between the two probability measures corresponding to the two hypotheses exceeds a threshold. Although there is no obvious notion of Radon-Nikodym derivative between two composite hypotheses, [44] demonstrates that a likelihood ratio test continues to be optimal for testing between composite hypotheses under some conditions on the games $u$ and $v$.

*Translation 10.* For concave infinite games $u$ and $v$, consider the composite hypotheses $\mathcal{P}_u$ and $\mathcal{P}_v$ that are their cores. Then, a minimax Neyman-Pearson test between the $\mathcal{P}_u$ and $\mathcal{P}_v$ can be constructed as the likelihood ratio test between an element of $\mathcal{P}_u$ and one of $\mathcal{P}_v$; in this case, the representative elements minimize the Kullback divergence between the two families.

In a certain sense, a converse statement can also be shown to hold. We refer to Huber and Strassen [44] for proofs and to Veeravalli et al. [45] for context, further results, and applications. Related results for minimax linear smoothing and rate distortion theory on classes of sources were given by Poor [46, 47] and to channel coding with model uncertainty were given by Geraniotis [48, 49].

## 8. DISCUSSION

The general approach to using cooperative game theory to understand rate or capacity regions involves the following steps. (i) Formulate the region of interest as the aspiration set of a cooperative game. This is frequently the right kind of formulation for multiuser problems. (ii) Study the properties of the value function of the game, starting with checking if it is balanced, if it is exact, if it has a large core, and ultimately by checking convexity or concavity. (iii) Interpret the properties of the game that follow from the discovered properties of the value function. For instance, balancedness implies a nonempty core, while convexity implies a host of results, including nice properties of the Shapley value. These are structural results, and their game-theoretic interpretation has the potential to provide some additional intuition.

There are numerous other papers which make use of cooperative game theory in communications, although with different emphases and applications in mind. See, for example, van den Nouweland et al. [50], Han and Poor [51], Jiang and Baras [52], and Yaïche et al. [53]. However, we have pointed out a very fundamental connection between the two fields—arising from the fact that rate and capacity regions are often closely related to the aspiration sets of cooperative games. In several exemplary scenarios, both classical and relatively new, we have reinterpreted known results in terms of game-theoretic intuition and also pointed out a number of open problems. We expect that the cooperative game theoretic point of view will find utility in other scenarios in network information theory, distributed inference, and robust statistics.

## ACKNOWLEDGMENTS

I am indebted to Rajesh Sundaresan for a detailed discussion that clarified my understanding of some of the literature, and the identification of an error in the first version of this paper. I am grateful to Sekhar Tatikonda and Edmund Yeh for useful conversations and help with references, and to A. R. Barron, A. M. Kagan, P. Tetali, and T. Yu for being collaborators on related work. Part of this work was done while I was a Visiting Fellow in the School of Technology and Computer Science at the Tata Institute of Fundamental Research, Mumbai, India, and I thank Vivek Borkar in particular for providing a stimulating environment.

**REFERENCES**

[1] R. J. La and V. Anantharam, "A game-theoretic look at the Gaussian multiaccess channel," in *DIMACS Series in Discrete Mathematics and Theoretical Computer Science*, 2002.

[2] G. Owen, *Game Theory*, Academic Press, Boston, Mass, USA, 3rd edition, 2001.

[3] O. N. Bondareva, "Some applications of the methods of linear programming to the theory of cooperative games," *Problemy Kibernetiki*, vol. 10, pp. 119–139, 1963, (Russian).

[4] L. S. Shapley, "On balanced sets and cores," *Naval Research Logistics Quarterly*, vol. 14, no. 4, pp. 453–560, 1967.

[5] L. S. Shapley, "Cores of convex games," *International Journal of Game Theory*, vol. 1, no. 1, pp. 11–26, 1971.

[6] M. Madiman and P. Tetali, "Information inequalities for joint distributions, with interpretations and applications," Tentatively accepted to *IEEE Transactions on Information Theory*, 2008.

[7] M. Maschler, B. Peleg, and L. S. Shapley, "The kernel and bargaining set for convex games," *International Journal of Game Theory*, vol. 2, pp. 73–93, 1972.

[8] J. Edmonds, "Submodular functions, matroids and certain polyhedra," in *Proceedings of the Calgary International Conference on Combinatorial Structures and Their Applications*, Gordon and Breach, Calgary, Canada, June 1969.

[9] R. J. Weber, "Probabilistic values for games," in *The Shapley Value*, pp. 101–119, Cambridge University Press, Cambridge, UK, 1988.

[10] T. Ichiishi, "Super-modularity: applications to convex games and to the greedy algorithm for LP," *Journal of Economic Theory*, vol. 25, no. 2, pp. 283–286, 1981.

[11] L. S. Shapley, "A value for n-person games," *Annals of Mathematics Study*, vol. 28, pp. 307–317, 1953.

[12] W. W. Sharkey, "Cooperative games with large cores," *International Journal of Game Theory*, vol. 11, no. 3-4, pp. 175–182, 1982.

[13] A. K. Biswas, T. Parthasarathy, J. A. M. Potters, and M. Voorneveld, "Large cores and exactness," *Games and Economic Behavior*, vol. 28, no. 1, pp. 1–12, 1999.

[14] D. Schmeidler, "Cores of exact games, I," *Journal of Mathematical Analysis and Applications*, vol. 40, no. 1, pp. 214–225, 1972.

[15] M. A. Rabie, "A note on the exact games," *International Journal of Game Theory*, vol. 10, no. 3-4, pp. 131–132, 1981.

[16] B. Lehmann, D. Lehmann, and N. Nisan, "Combinatorial auctions with decreasing marginal utilities," in *Proceedings of the 3rd ACM Conference on Electronic Commerce (EC '01)*, pp. 18–28, Tampa, Fla, USA, October 2001.

[17] U. Feige, "On maximizing welfare when utility functions are subadditive," in *Proceedings of the 38th Annual ACM Symposium on Theory of Computing (STOC '06)*, pp. 41–50, Seattle, Wash, USA, May 2006.

[18] D. Slepian and J. K. Wolf, "Noiseless coding of correlated information sources," *IEEE Transactions on Information Theory*, vol. 19, no. 4, pp. 471–480, 1973.

[19] T. M. Cover, "A proof of the data compression theorem of Slepian and Wolf for ergodic sources," *IEEE Transactions on Information Theory*, vol. 21, no. 2, pp. 226–228, 1975.

[20] S. Fujishige, "Polymatroidal dependence structure of a set of random variables," *Information and Control*, vol. 39, no. 1, pp. 55–72, 1978.

[21] T. P. Coleman, A. H. Lee, M. Médard, and M. Effros, "Low-complexity approaches to Slepian-Wolf near-lossless distributed data compression," *IEEE Transactions on Information Theory*, vol. 52, no. 8, pp. 3546–3561, 2006.

[22] R. Cristescu, B. Beferull-Lozano, and M. Vetterli, "Networked Slepian-Wolf: theory, algorithms, and scaling laws," *IEEE Transactions on Information Theory*, vol. 51, no. 12, pp. 4057–4073, 2005.

[23] A. Ramamoorthy, "Minimum cost distributed source coding over a network," in *Proceedings of IEEE International Symposium on Information Theory (ISIT '07)*, Nice, France, June 2007.

[24] B. Rimoldi and R. Urbanke, "A rate-splitting approach to the Gaussian multiple-access channel," *IEEE Transactions on Information Theory*, vol. 42, no. 2, pp. 364–375, 1996.

[25] R. Ahlswede, "Multi-way communication channels," in *Proceedings of the 2nd International Symposium on Information Theory (ISIT '71)*, Hungarian Academy of Sciences, Tsahkadsor, Armenia, September 1971.

[26] H. Liao, "A coding theorem for multiple access communication," in *Proceedings of IEEE International Symposium on Information Theory (ISIT '72)*, Asilomar, Calif, USA, January 1972.

[27] D. Slepian and J. K. Wolf, "A coding theorem for multiple access channels with correlated sources," *Bell System Technical Journal*, vol. 52, no. 7, pp. 1037–1076, 1973.

[28] T. S. Han, "The capacity region of general multiple-access channel with certain correlated sources," *Information and Control*, vol. 40, no. 1, pp. 37–60, 1979.

[29] T. M. Cover and J. A. Thomas, *Elements of Information Theory*, John Wiley & Sons, New York, NY, USA, 1991.

[30] E. M. Yeh, *Multiaccess and fading in communication networks*, Ph.D. thesis, Massachusetts Institute of Technology, Cambridge, Mass, USA, 2001.

[31] D. N. C. Tse and S. V. Hanly, "Multiaccess fading channels. I. Polymatroid structure, optimal resource allocation and throughput capacities," *IEEE Transactions on Information Theory*, vol. 44, no. 7, pp. 2796–2815, 1998.

[32] A. Lapidoth and S. Shamai, "The Poisson multiple-access channel," *IEEE Transactions on Information Theory*, vol. 44, no. 2, pp. 488–501, 1998.

[33] S. C. Tatikonda, *Control under communication constraints*, Ph.D. thesis, Massachusetts Institute of Technology, Cambridge, Mass, USA, 2000.

[34] M. Madiman, A. R. Barron, A. M. Kagan, and T. Yu, "Minimax risks for distributed estimation of the background in a field of noise sources," in *Proceedings of the 2nd International Workshop on Information Theory for Sensor Networks (WITS '08)*, Santorini Island, Greece, June 2008, preprint.

[35] M. A. Girshick and L. J. Savage, "Bayes and minimax estimates for quadratic loss functions," in *Proceedings of the 2nd Berkeley Symposium on Mathematical Statistics and Probability*, pp. 53–73, University of California Press, Berkeley, Calif, USA, July-August 1951.

[36] C. E. Shannon, "A mathematical theory of communication," *Bell System Technical Journal*, vol. 27, pp. 379–423, 1948.

[37] P. Bergmans, "A simple converse for broadcast channels with additive white Gaussian noise," *IEEE Transactions on Information Theory*, vol. 20, no. 2, pp. 279–280, 1974.

[38] L. Ozarow, "On a source-coding problem with two channels and three receivers," *Bell System Technical Journal*, vol. 59, no. 10, pp. 1909–1921, 1980.

[39] M. Costa, "On the Gaussian interference channel," *IEEE Transactions on Information Theory*, vol. 31, no. 5, pp. 607–615, 1985.

[40] Y. Oohama, "The rate-distortion function for the quadratic Gaussian CEO problem," *IEEE Transactions on Information Theory*, vol. 44, no. 3, pp. 1057–1070, 1998.

[41] M. Madiman and A. Barron, "Generalized entropy power inequalities and monotonicity properties of information," *IEEE Transactions on Information Theory*, vol. 53, no. 7, pp. 2317–2329, 2007.

[42] M. Madiman, "On the entropy of sums," in *Proceedings of IEEE Information Theory Workshop (ITW '08)*, Porto, Portugal, May 2008.

[43] G. Choquet, "Theory of capacities," *Annales de L'institut Fourier*, vol. 5, pp. 131–295, 1954.

[44] P. J. Huber and V. Strassen, "Minimax tests and the Neyman-Pearson lemma for capacities," *Annals of Statistics*, vol. 1, no. 2, pp. 251–263, 1973.

[45] V. V. Veeravalli, T. Basar, and H. V. Poor, "Minimax robust decentralized detection," *IEEE Transactions on Information Theory*, vol. 40, no. 1, pp. 35–40, 1994.

[46] H. V. Poor, "The rate-distortion function on classes of sources determined by spectral capacities," *IEEE Transactions on Information Theory*, vol. 28, no. 1, pp. 19–26, 1982.

[47] H. V. Poor, "Minimax linear smoothing for capacities," *Annals of Probability*, vol. 10, no. 2, pp. 504–507, 1982.

[48] E. Geraniotis, "Minimax robust coding for channels with uncertainty statistics," *IEEE Transactions on Information Theory*, vol. 31, no. 6, pp. 802–811, 1985.

[49] E. Geraniotis, "Robust coding for multiple-access channels," *IEEE Transactions on Information Theory*, vol. 32, no. 4, pp. 550–560, 1986.

[50] A. van den Nouweland, P. Borm, W. van Golstein Brouwers, R. Groot Bruinderink, and S. Tijs, "A game theoretic approach to problems in telecommunication," *Management Science*, vol. 42, no. 2, pp. 294–303, 1996.

[51] Z. Han and H. V. Poor, "Coalition games with cooperative transmission: a cure for the curse of boundary nodes in selfish packet-forwarding wireless networks," in *Proceedings of the 5th International Symposium on Modeling and Optimization in Mobile, Ad Hoc,and Wireless Networks (WiOpt '07)*, Limassol, Cyprus, April 2007.

[52] T. Jiang and J. S. Baras, "Fundamental tradeoffs and constrained coalitional games in autonomic wireless networks," in *Proceedings of the 5th International Symposium on Modeling and Optimization in Mobile, Ad Hoc,and Wireless Networks (WiOpt '07)*, Limassol, Cyprus, April 2007.

[53] H. Yaïche, R. R. Mazumdar, and C. Rosenberg, "A game theoretic framework for bandwidth allocation and pricing in broadband networks," *IEEE/ACM Transactions on Networking*, vol. 8, no. 5, pp. 667–678, 2000.